# PWCT: Visual Language for IoT and Cloud Computing Applications and Systems

Mahmoud S. Fayed, Muhammad Al-Qurishi, Atif Alamri, Ahmad A. Al-Daraiseh
Research Chair of Pervasive and Mobile Computing
College of Computer and Information Sciences
King Saud University, Riyadh, Saudi Arabia
msfclipper@yahoo.com,{qurishi, atif, adaraiseh}@ksu.edu.sa

## ABSTRACT

Developing IoT, Data Computing and Cloud Computing software requires different programming skills and different programming languages. This cause a problem for many companies and researchers that need to hires many programmers to develop a complete solution. The problem is related directly to the financial cost and the development time which are very important factors to many research projects. In this paper we present and propose the PWCT visual programming tool for developing IoT, Data Computing and Cloud Computing Applications and Systems without writing textual code directly. Using PWCT increase productivity and provide researchers with one visual programming tool to develop different solutions.

## INTRODUCTION

The requirements for software applications are increasing because computers are now very large part of our everyday life. Software programs are more complex because of many factors. Nowadays programs run on a wide variety of hardware like high-performance clusters, personal computers, embedded devices and distributed systems. Applications are developed for different fields and the cost can vary significantly as free open-source software compete with proprietary software. Decreasing cost, improving reliability and increasing scalability are some of the requirements facing software developers. In the age of information technology, software development plays a vital role to respond to researchers, companies and organizations need of high quality information systems. This leads to the need to more programmers and more productive software development tools to be able to respond quickly to companies need with high quality [1].

Towards solving this problem, In this article we propose a novel visual programing language called PWCT. The remainder of this paper is organized as follows. Section 2 describes related works; Section 3 illustrates the software architecture. Section 4 presents evaluation for our proposed method, whereas section 5 shows the current real users and applications. Section 6 presents the future work. Finally concluding remarks are made in Section 7.

## RELATED WORD

After the success of many domain specific visual programming languages (VPLs) like Scratch, Alice by reaching millions of users worldwide, it's expected to find more interest in creating new VPLs that help novice programmers to learn programming and help mainstream programmers to create high-quality programs faster. Such languages must be designed carefully to solve current VPLs' issues without creating new problems and this is an important factor for new VPLs to gain popularity. Also, to increase the usage of these new VPLs, they need to compete and/or integrate with mature development tools that are based on popular programming languages such as C++, Java, C#, Python and Ruby [1-5].

A Visual Programming Language (VPL) combines the features of integrated development environments and information systems to provide a visual tool for applications' development. In addition to that, VPLs enable the development of applications and computer programs using more than one dimension, and provide a programming system based on interaction with graphical elements that mix between Text, Shapes, Colors, and Time, instead of typing textual source code [1-3, 8-10].

Currently, there are many VPLs in the market, but most of the successful and widely used ones are educational tools such as: Scratch, Alice & Kodu [6]; or domain-specific such as: Max/MSP (Music and Multimedia) and LabView (Data Acquisition, Instrument Control and Industrial Automation) [12]. General-purpose VPLs like Limnor, Forms/3, Tersus, and Envision [1,7,13] exist as well, but these languages are not widely used, According to TIOBE Index that measures the popularity of programming languages [1, 11].

Scratch is developed at the Massachusetts Institute of Technology Media Laboratory. It is a new VPL and environment that supports the creation of interactive stories, games, animations, music and art projects (as shown in figure 1). Scratch allows educators to reduce the cognitive load that learners experience when they are initially introduced to programming.

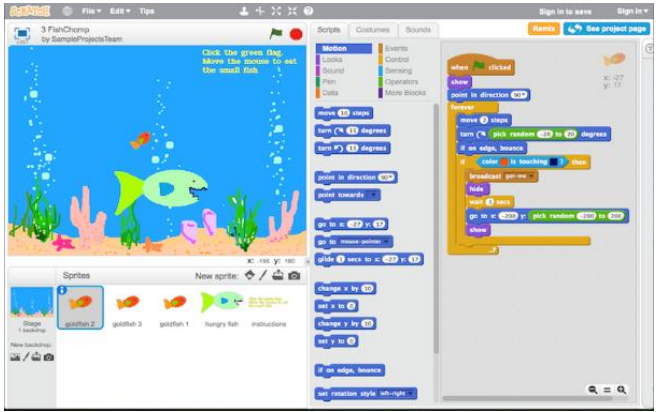

**Figure 1. Scratch Environment**

Alice is a VPL where objects are manipulated in a 3D world (as shown in figure 2). It was developed at Carnegie Mellon University; it gives students the chance to learn about object-oriented programming concepts without the syntax frustrations imposed by text-based programming languages. With Alice, a programmer using a Graphical User Interface (GUI) environment selects program constructs and methods from lists of available choices.

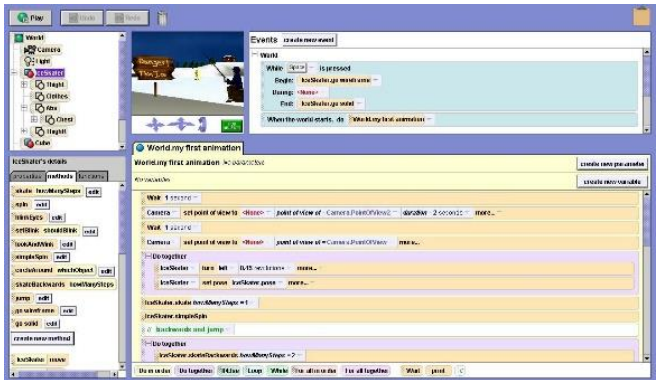

**Figure 2. Alice Environment**

Tersus is an open source VPL developed by Tersus Software Ltd. It is used to build rich web and mobile applications by visually defining user interface, client side behavior, and server side processing (as shown in figure 3). It is a general purpose language that utilizes flow diagrams.

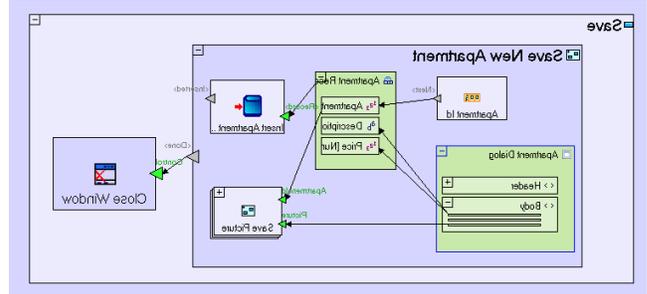

**Figure 3. Tersus – Data Flow Diagram**

Limnor (shown in figure 4) developed by Longflow Enterprises Limited is general-purpose VPL. It is based on visual studio .NET. It can be used to create most applications, interactive multimedia kiosks, sales presentations, database applications with interactive query and search, business management systems, internet payphone kiosks, etc. Non-technically oriented users can use it [7]. Limnor supports code generation only in C#.

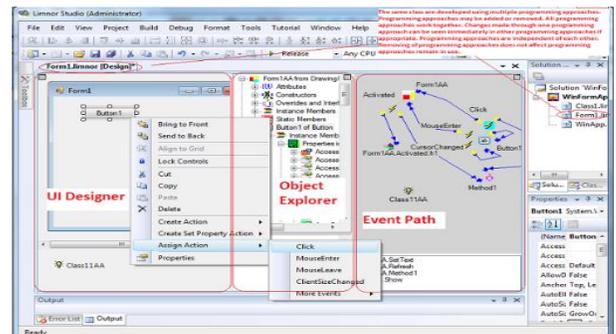

**Figure 4. Limnor Environment**

## SOFTWARE ARCHITECTURE

PWCT is composed of three main layers: VPL Abstract Layer; Middleware Layer; and System Layer. Figure 5. Shows each layer and its constituent components within PWCT's architecture.

### 3.1. The VPL Layer

The VPL layer is the presentation layer. It provides all the functionality needed to allow a user to perform a specific task. It is composed of four important components: the Goal/Module Designer; the Components Browser; the Interaction Pages; and the Form Designer. A user would interact with at least three components to produce a meaningful program. For example, a user starts with the Goal/Module Designer to create a new goal (in other words module). Next, the user might choose a visual component from the Components Browser. Then, the user might work with the Interaction Pages to provide the needed data or to specify some attributes. If the user wants to develop a GUI, they can use the Form Designer to do so.

## 3.2. The Middleware Layer

PWCT's middleware connects the user's view (i.e. the four components in the VPL Layer) with the system's view (i.e. processes in the System Layer).

This layer has four main functionalities: interpretation of user's interactions with components in the VPL Layer; generating the Steps Tree according to user's interactions; code matching and code masking for the selected language to generate source code files; and generating the source code.

As demonstrated in figure 5, the Middleware layer consists of three units; the Step Generator, a database, and the source code files. The most important part of the middleware layer is the Step Generator. Generating the code in this layer is fairly simple and straight forward.

## 3.3. The System Layer

This layer is a low-level layer that deals with the source code generated by the upper layer. PWCT generates and executes code in four text-based programming languages namely C, C#, Harbour, and Python and hence, the System Layer had to support the different mechanisms in all four languages to produce executable code.

Programs that were written by text-based languages are implemented either by an interpreter, a compiler or a combination of both. The System Layer support compilers and interpreters at the same time. After generating the source code in any of the supported text-based language (C, C#, Harbour, and Python), executing the code follows the procedures of that specific language. PWCT doesn't impose any restrictions on the process of building the source code.

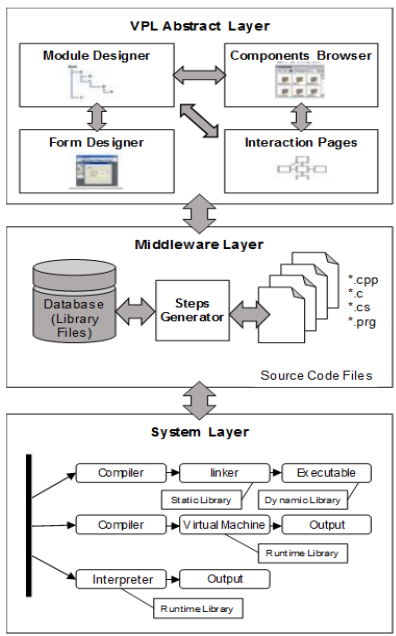

**Figure 5. PWCT Architecture**

## EFFICIENCY EVALUATION

In this experiment PWCT competed against two well-known general-purpose VPLs (Tersus and Limnor), and two reputable IDEs (Visual Studio and Net Beans). The selection of these VPLs was based on selecting tools that are (General Purpose, Ready for practical use, Support code generation in popular textual languages and Used already by developers in developing business applications). The selection of the IDEs is based on selecting popular IDEs for popular textual programming languages like Visual Studio IDE for the C# language and Netbeans IDE for the Java language. There are many other tools that are excluded because they don't satisfy the criteria or because they are very similar to the selected tools.

One professional programmer for each language was hired to develop programs in that specific language. All five programmers were asked to develop 20 simple to intermediate applications. The authors monitored the time and memory needed to complete each task. Table 1 shows the averages of all tasks. It is extremely obvious that PWCT's outperformed the other four languages by a considerable margin. PWCT's time was 8.5% less than Visual Studio (the second best) and 55.6% better than Net Beans. This criterion is very important as it hints at the productivity of the language. Similarly, while PWCT used only 4.52 MB of RAM on average, the second best (Tersus) used 12.124 MB. Figure. 6 shows that PWCT outperforms all other languages as far as development time and memory are concerned.

**Table 1. Comparison between PWCT, Limnor, Tersus, Visual Studio and Netbeans**

| Tool | Type | Development Time (Seconds) | Memory (MByte) | User Steps |
|---|---|---|---|---|
| PWCT | VPL | 106.66 | 4.52 | 41 |
| Limnor | VPL | 144 | 15.8 | 54 |
| Tersus | VPL | 140 | 12.124 | 46 |
| Visual Studio | IDE | 115.66 | 58.144 | 26 |
| NetBeans | IDE | 165 | 17.3 | 33 |

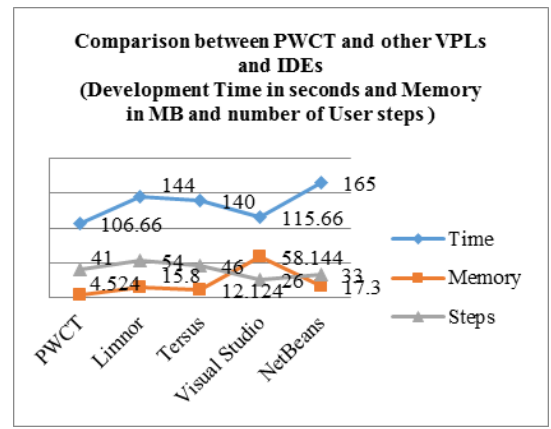

**Figure 6. Comparison between PWCT and other tools**

## REPUTATION EVALUATION

PWCT is very popular on Sourceforge [14]. Out of the 300,000 projects on Sourceforge, PWCT ranks third in the science and engineering category as of April 2016. The number of weekly downloads (environment, samples and tutorials) according to Sourceforge is more than 70,000 downloads.

According to Sourceforge's information, USA has the highest download rate of all countries followed by the Italy, India, United Kingdom, France, Germany, Algeria and Canada in the 8th place. Moreover, according to Sourceforge's user's satisfaction information, 93% of all users are satisfied with PWCT while only 7% are not as illustrated in Figure 7.

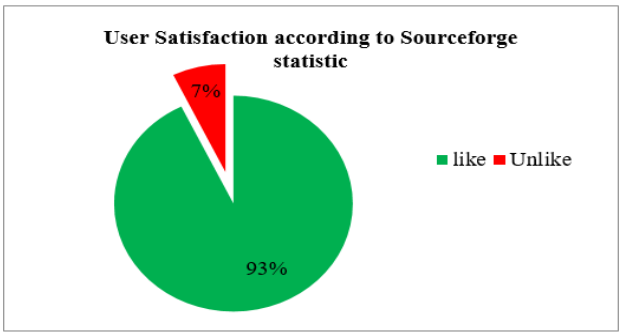

**Figure 7. User Satisfaction**

Many systems and applications were developed using PWCT in several domains, such as: multi-media, database, network programming, business management systems, IoT, Cloud Computing and Mathematics. PWCT has not ignored the business management systems, where a group of programmers in Riyadh Valley Company, built a business application for the company to follow-up its projects and monitor the performance of its divisions as showing in the dashboard in Fig. 8. The screen snapshot in Fig.9 depicts a multi-user client-server database application for managing car-hiring companies. This software achieved by one of the active users of PWCT in Netherlands. The screen snapshot in Figure 10. Depicts an Implementation of a new Localized Algorithm for detection of Critical Nodes using PWCT [15,16]. The application is developed by a group of researchers in King Saud University. Also some programming languages are developed using PWCT like the Supernova programming language and the Ring programming language [17].

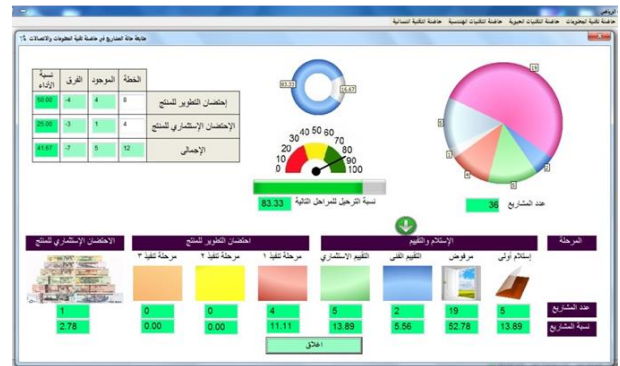

**Figure 8. RVC application developed using PWCT**

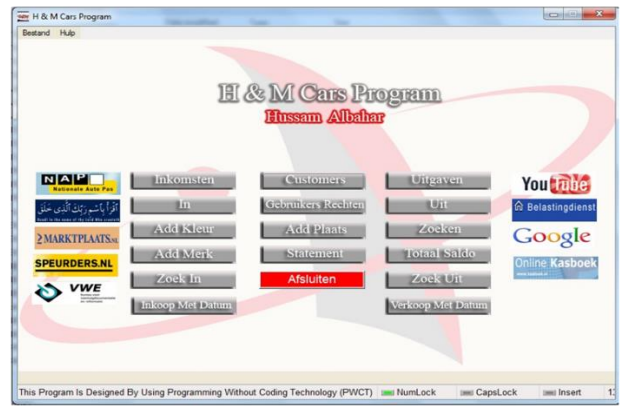

**Figure 9. Car-hiring application developed using PWCT**

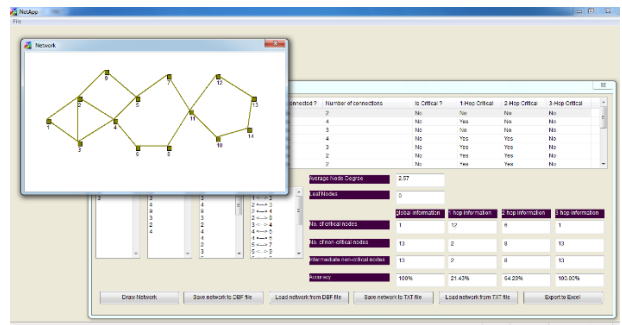

**Figure 10. Critical Nodes application developed using PWCT**

## FUTURE WORK

A more advanced version of PWCT is under development. It will support more platforms and will provide code generation in more textual languages. The components will be redesigned to provide more productivity. An interesting feature will be the support of our new textual programming language that we designed for the development of PWCT 2.0. This language called Ring. Using the Ring language we will have behind the Visual Programming Language, a scripting language that uses Natural Language Programming and Declarative Programming Paradigms. This will provide high-level of abstraction when the developer decide to switch from visual programming and look at the generated source code. Also the other textual programming languages will still be supported when the developer select them.

An interesting domain of visual programming languages is their usage in Education. Tools like Scratch and Alice exist and are very successful. We will provide courses about PWCT through distance-learning to students and developers to study the learning curve and the time required for mastering the tool in applications and systems development. Also we will work on developing more projects about IoT, Data Computing and Cloud Computing applications and systems that integrate between all of these domains together.

## CONCLUSION

The PWCT visual programming language is presented. PWCT is designed for IoT, Data Computing and Cloud Computing Applications and Systems. PWCT was designed to compete with textual programming languages such as C, C++ and Java. The novelty of this work comes from the ability to use PWCT to create large applications and systems like textual programming languages compilers and virtual machines, IoT, Data Computing and Cloud Computing Applications. PWCT uses graphical components to replace textual code in an easy and seamless process. PWCT's architecture and design were covered in this paper. Moreover, data from Sourceforge indicates that PWCT is a popular, well accepted, and highly downloadable VPL.

The development of many business applications and some programming languages like Supernova and Ring using PWCT demonstrates that the new VPL can be used in serious development for creating many real world projects.

## ACKNOWLEDGMENTS

We would like to say thanks to the PWCT Team all over the world for contributing to the software in testing and marketing. The project was supported by King Saud University, Deanship of Scientific Research, Research Chair of Pervasive and Mobile Computing.